# USING SENSORS IN THE WEB CRAWLING PROCESS

Ilya Zemskov
*Omsk State University, Department of Computer Science, Division of Cybernetics*
*Russia, 644077, Omsk, pr. Mira, 55-A*
*zemskov@univer.omsk.su*

**ABSTRACT**

This paper offers a short description of an Internet information field monitoring system, which places a special module-sensor on the side of the Web-server to detect changes in information resources and subsequently reindexes only the resources signaled by the corresponding sensor. Concise results of simulation research and an implementation attempt of the given "sensors" concept are provided.

## 1. INTRODUCTION

With a certain degree of abstraction search engines can be said to be, practically, monitoring the information field of Internet. Here and further on monitoring will be understood as a process of representing the content of available information resources of different media-types. There have been attempts (Brandman 2000) in the development history of the information field Internet monitoring technologies to impart a more active role in the process of monitoring to server software in order to decrease the calling rate of the communication channels of the information resources owners. But it can be observed that at present the monitoring process is still run only with the help of special robot-programs, which are the main source of additional calling rate and, moreover, are limited in size and depth (Raghavan 2001) of the encompassed area of the resources being monitored.

Analyzing existing publications pertaining to the problems monitoring the information field of Internet, two concepts of creating monitoring systems can be identified (such systems often appear to be only parts of larger search engines). The first one is the most popular and widespread concept – the "robot" concept. The second one is the concept of "mobile robots" (Fiedler 1999). Our goal and the goal of this article in particular is an attempt to offer for the consideration of international scientific community a third concept, namely the "sensors" concept. For this purpose general ideas of the "sensors" concept and some results of our simulation experiments will be briefly described and the course of our work in creating a model version of software system, which will implement the suggested ideas, will be produced.

## 2. «SENSORS» CONCEPT

As it has already been mentioned, the creators of Internet monitoring systems often pay their attention to the server software in order to expand it with such features, which will increase the efficiency of the monitoring process. The "sensors" concept, at its core, also offers to expand the server software (in this research we limit ourselves to the consideration of Web-server software, but it is obvious that these ideas can be applied to any other types of servers) with a special software-based module or "sensors".

The description of how the "sensors" algorithm works can be started by mentioning the following facts. First of all, Web-server appoints to a received request a certain sequence of bytes (this sequence has a unique value, for example, MD5 checksum), which is the demanded information. Secondly, to each response the server appoints a certain set of characteristics (URL, size, content type, status code, etc), which give an idea of the received response. Now it becomes obvious that the main function of the passive part of the server will be keeping a special database. This "database" should contain the information about the received request and the data, characterizing the response given by the Web-server. The active part of the sensor had the function of tracking changes in responses to the same requests and informing the indexing robot of the search engine about the changes, that have taken place (in the minimum variant it is enough to make a GET request with the transmission of the wanted parameters, but it is necessary for the robot to be ready to receive such requests). For its part the robot decides for itself what to do with the received notification.

Thus, on each Web-server a special module needs to be installed and adjusted for effective work (at the minimum a range of requests forbidden for indexation similarly to the file robots.txt should be indicated). It stands to reason that it will allow to optimize the calling rate that the monitoring system places on the monitored environment and thus reduce the inefficient financial costs for both sides taking part in the monitoring process (the owners of information resources and the owners of search engine).

## 3. SIMULATION

For a preliminary analysis of the peculiarities of functioning of the monitoring system built with the use of the "sensors" concept we decided to conduct a simulation research. On the basis of common sense and the analysis of existing publications we have chosen current freshness (%) of the monitoring system index and the amount (bytes) of information downloaded by the robot as the main performance criteria of the monitoring system. To receive a numerical value of the given criteria we have designed a model and a simulation software system. For the purpose of objective evaluation of the received data we have also designed a model of the "robot" concept and have expanded the existing software system with a corresponding simulation program. The improvement of the software system and the simulation experiments are still continuing, but even now we can describe the results obtained at the first stage.

At the first stage a plan of the experiments, which consisted of 9 series (3 runnings each), was worked out. In each series we modeled the functioning of a system monitoring the information status of 200000 information resources (html-pages are meant) during 10 days (8640000 units) of modeling time. As 1 unit of modeling time we recognized 10 milliseconds (msec) of real time. The initial size (bytes) of every information resource is determined using uniform distribution with the minimum value equal to 65 bytes (minimal size of an html-markup of a page) and maximum possible value equal to 122880 bytes (recommended maximum size for a page). Downloading time of any information resource is determined by uniform distribution with the minimum value equal to 1 unit of modeling time and the maximum value equal to 40 units of modeling time.

The time of receiving requests for downloading every information resource and the time of occurrence of a change in status or in content of the information resource is distributed exponentially. The corresponding rates can be found in Table 1. They are marked through a slash: changes /requests. For example, for the series [2-3] the rate of changes equals 10, and the rate of requests equals 50 (i.e. during 10 days the resource changes its status 10 times and is requested by users 50 times).

Table 1. Rates

|   | 1     | 2     | 3       |
|---|-------|-------|---------|
| 1 | 1 / 1 | 5 / 1 | 10 / 1  |
| 2 | 1 / 50| 5 / 50| 10 / 50 |
| 3 | 1 / 100| 5 / 100| 10 / 100|

In the model 6 types of possible status-changes of the information resource are determined: status code 403, status code 404, status code 500, decreasing in size of the page, increasing in size of the page, page available (no change).

When modeling a system built according to the "robot" concept, only one sample robot is doing the monitoring and he has to "make a round of" the whole set of information resources. But while modeling a system built according to "sensors" concept, it's the sensors who, do the monitoring and send a notification signal to the robot. The arrival time of the notification from a sensor of any information resource is determined by uniform distribution with the minimum value equal to 1 unit modeling time and the maximum value equal to 3 units modeling time. On receiving a notification signal the robot initiates downloading of the changed page regardless of the number of active downloads initiated in response to signals received earlier.

The modeling software system was designed with the help of Python 2.2, modeling library SimPy 1.0 (Simula-like), library MySQLdb for access to DB MySQL 3.23.38. The experiments are run on Pentium 4, 512Mb, 60Gb HDD, Windows 2000 Professional.

Figure 1. Freshness (%) of index: a – "robot" concept, b – "sensors" concept

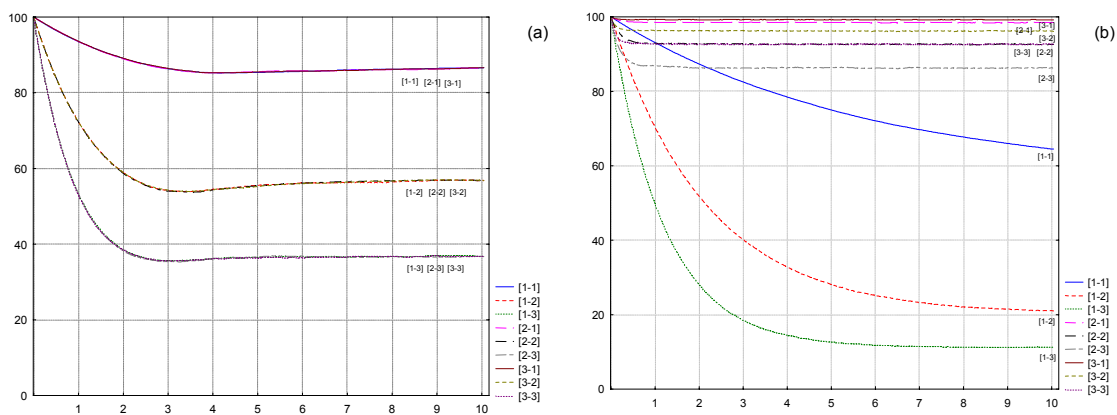

Figure 2. Amount (Gb) of downloaded information: a – "robot" concept, b – "sensors" concept

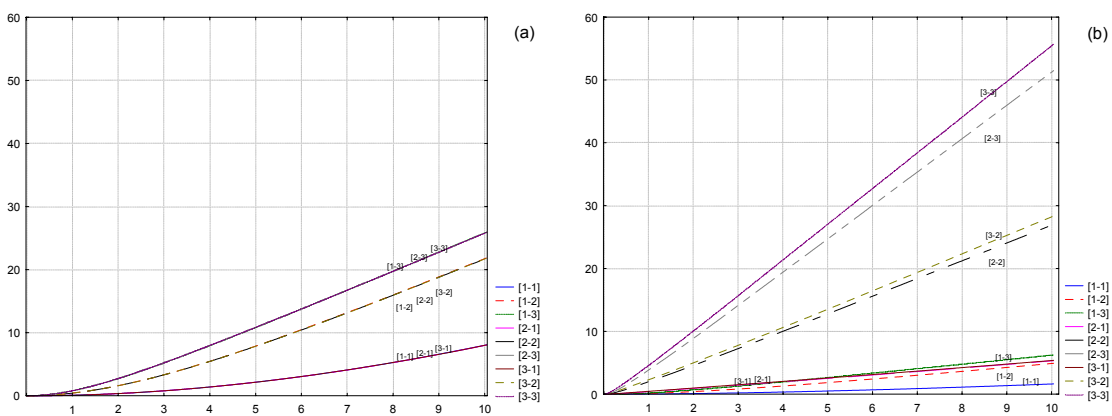

Figures 1 and 2 show the graphs built on the basis of the data, which was received as the result of the experiments. Measurements of the corresponding characteristics were made every 10000 units of modeling time. Carrying out of the experiments took a month of real time.

## 4. PRACTICAL IMPLEMENTATION

To assure ourselves in feasibility of the ideas suggested by the "sensors" concept we decided to undertake an attempt to create a corresponding software system. This software system will obviously have to include modules, which will carry out the functions of the sensor, a special indexing robot and a search interface. It needs to be mentioned that the creation of a search interface is not a part of this research, that is why it became important to find a ready-made search engine with well developed functionality (alternative encoding support is a required) and a possibility to make necessary modifications in the source code. Due to insufficient financing of this research successful commercial products were unavailable to us and we turned to software products with open source code. At this point in our research we are inclined to use «Fluid Dynamics Search Engine» (Perl source code is available).

The analysis of Web-server platforms (Apache, IIS, Lotus Domino) used in Omsk State University showed that they all have as their part special API for creating modules expanding functional capabilities.

Yet at the first stage of this work we decided to limit ourselves to designing a sensor module only for the Web-server Apache, as the possibilities of mod_perl environment fit well with our goals.

The test results of the first versions of sensor modules showed that our ideas are perfectly viable. Therefore this work will be continued and its result will be a search engine covering all Web-servers of our university.

## 5. CONCLUSION

It stands to reason that the main objection to the suggested concept will be unwillingness of the owners of information resources to place additional modules on their Web-servers, especially the modules, which "spy". There are several approaches to this problem. The first one is the simulation research, which will demonstrate the substantial economic advantage gained by implementation of the new concept. The other approaches to this problem can be briefly characterized as approaches to the ways and methods of module distribution (appropriate license policy, availability of detailed documentation, including the module in the new versions of installation package for the Web-server software).

Another obvious objection will be the fact that for the implementation of the suggested concept in the context of the whole Internet and obtaining substantial reduction in financial costs for all members of the monitoring process a united center, which will receive notifications about any changes, will have to be created. The solution of this problem may lie in creating an independent noncommercial monitoring center on the analogy of ICANN.